# Cybernetic modeling of Industrial Control Systems: Towards threat analysis of critical infrastructure

Abhinav Biswas
Technical Officer, IT Services Division,
Electronics Corporation of India Limited, ECIL
Hyderabad, India
abhinavbiswas@ecil.gov.in

Sukanya Karunakaran
Technical Officer, Corporate Research & Development,
Electronics Corporation of India Limited, ECIL
Hyderabad, India
sukanya_karun@ecil.co.in

*Abstract* — Industrial Control Systems (ICS) encompassing resources for process automation are subjected to a wide variety of security threats. The threat landscape is arising due to increased adoption of Commercial-of-the-shelf (COTS) products as well as the convergence of Internet and legacy systems. Prevalent security approaches for protection of critical infrastructure are scattered among various subsystems and modules of ICS networks. This demands a new state-of-the-art cybernetic model of ICS networks, which can help in threat analysis by providing a comprehensive view of the relationships and interactions between the subsystems. Towards this direction, the principles of the Viable System Model (VSM) are applied to introduce a conceptual recursive model of secure ICS networks that can drive cyber security decisions.

## I. INTRODUCTION

Industrial Control Systems (ICS) are an integral part of a nation's critical infrastructure providing requisite resources for process control automation of critical entities like power grids, emergency communications systems, air traffic control networks, nuclear power plants etc. Control & Instrumentation (C&I) systems of critical infrastructure are so vital that the potential disruption of such entities would have a debilitating impact on national security, industrial economy, public safety etc. and hence need to be protected from deliberate sabotage attacks. Historically, C&I systems were closed and often built on specialized proprietary protocols and interfaces. However, nowadays C&I systems are adopting components from open Commercial-of-the-shelf (COTS) technology for increased flexibility and cross platform compatibility. The old esoteric systems are being replaced with new open products that have connectivity to corporate intranets and, in some cases, the public Internet. However, this paradigm shift in adopting openness and Internet connectivity is achieved at the cost of decreased security against digital intrusion. Modern embedded systems are riddled with vulnerabilities, which are subjected to potential cyber threats like state-sponsored external attacks as well as malicious insider threats. Hence, it has become a challenge for researchers to introduce requisite security and sufficient safety features in modern ICS networks to counter the newly posed threats. Furthermore, since the C&I systems typically have much longer life cycle than commercial IT systems, there is an increasing demand for a security assurance model of ICS networks that encompasses the entire running cycle of C&I systems.

Majority of the prevalent ICS security approaches are adaptations of conventional methods for securing traditional IT systems which address only a part of the threat landscape and are not specifically tailored for elements of ICS networks like sensors/actuators etc. Also in traditional ICS network designs the identification of threats and vulnerabilities is conducted in each domain of the ICS network separately and independently, omitting the interactions between the underlying field components. Therefore, a novel holistic approach that can overcome these drawbacks of past approaches is the need of the hour for ICS network security.

Hence, this article introduces a conceptual recursive model of a secure ICS network that can withstand the challenges of modern threats by adopting the principles of Viable System Model (VSM). Towards this direction, an intuitive 7-level secure network design model is proposed that provides a holistic view of the critical sections of ICS networks. The proposed model uses the cybernetic construct of VSM to methodologically emphasize the relationships and interactions between different entities within & outside the ICS Infrastructure system. The article demonstrates how a layered design principle can ease threat visualization & help increase security assurance at all levels.

More specifically, this article's key contributions are:

- A formal demonstration of the seven-layered cybernetic model of ICS networks based on VSM principles.

- Illustration of recursive relationships and interactions between the different C&I systems at each layer of VSM.

- Threat analysis of all layers regarding susceptibility to modern cyber attacks and how to combat them.

## II. BACKGROUND ANALYSIS

Government organizations that deal with critical infrastructure nowadays confront targeted-attacks like Advanced Persistent Threats (APT), which are destined for specific industrial sabotage. APTs are coordinated cyber attacks that target the specific industry over a persistent time period of months or even years and try to penetrate into the ICS network stealthily using advanced malware that exploit zero-day vulnerabilities. C&I systems comprise a plethora of loosely coupled COTS systems like Supervisory Control And Data Acquisition (SCADA) systems, Programmable Logic Controllers (PLCs), Remote Terminal Units (RTUs) etc. that demand for 24/7 remote access for engineering, operations or technical support and hence are jellied with the corporate network. As the modern dynamics of ICS networks is moving towards this sort of interconnected architecture, it is indirectly opening doors for the external attackers to penetrate. Further, since COTS technology is open and flexible, software driven C&I systems are now being subjected to hacking and targeted attacks like APTs much more than before.

Another contemporary threat faced by ICSs is that of Insider attacks. Insiders are confided members/operators who have administrative credentials to resources and knowledge about the network architecture. An insider with malicious intent can be very dangerous to the security of ICS networks. With the adaptation of novel technologies, threats to C&I systems may also arise due to negligent operational mistakes, inappropriate testing, wrongly defined security policy and even lack of knowledge about information security. These inherent threats if exploited, may not only partially paralyze the services but may introduce catastrophic failure of other critical infrastructure services due to cascading effects of the inter-related functional areas.

Ralph [1] explains the devastating success of Stuxnet malware that was used to infect the SCADA systems of Iran's nuclear facilities. Stuxnet contained a PLC rootkit that led the centrifuges to spin out of control and ultimate reactor meltdown. The proliferations of sophisticated Stuxnet-like malware such as Duqu, Flame, and Gauss in recent years reveal the imminence of modern threats and the shortcomings of our detection and response counter-measures. Hence, in order to address ICS security, it is indeed a necessity for organizations to adopt a defensive security framework that can maintain the confidentiality, integrity and availability (CIA) of information at all levels. The intrusions and risks can be mitigated only by a proactive defense-in-depth approach starting from the bottom level of field data acquisition to the top level of data sharing with the external world. Considering the fact that ICSs are often very long-lived installations and the duration of innovation cycles in ICSs is very different from those in conventional Information and Communication technology (ICT), this new level of demand for security assurance has necessitated the requirement to build up a secure ICS network model that is viable enough to generate effective deterrents and countermeasures against the known threats.

Critical Infrastructure security assurance and risk assessment methods are divided into two distinct categories viz. the sectorial approach referring to those which treat each sector separately and the flow approach that follows a systemic approach examining ICSs as interconnected networks. Most prevalent approaches [2] fall into the first category, which has limitations when it is applied in cross-sectorial environments. The resilience concerns are not addressed or exist only implicitly. Wang [3] proposed an excellent simulation environment for SCADA security analysis taking into account the interdependencies to an extent and provided resilience by focusing on network communications, but without considering cyber-threats that are not network-driven. The proposed model described in this article falls in the second category that utilizes VSM in a much more flow-oriented manner for identification of threats. The inherent scalable nature of the VSM ensures resilience whereas its systemic behavior allows the investigation of the critical components and emerging interconnections.

*The VSM paradigm*

Stafford Beer [4] originally designed the Viable System Model (VSM) to model the viability of an autonomous system. A viable system needs to meet the demands of survival in a changing environment and must be adaptable. The VSM expresses a cybernetic model for viable systems that is capable of autonomy and divides this system into three fundamental parts viz. Management, Operations and Environment as shown in Fig. 1a. The Operations part involves all the jobs that take

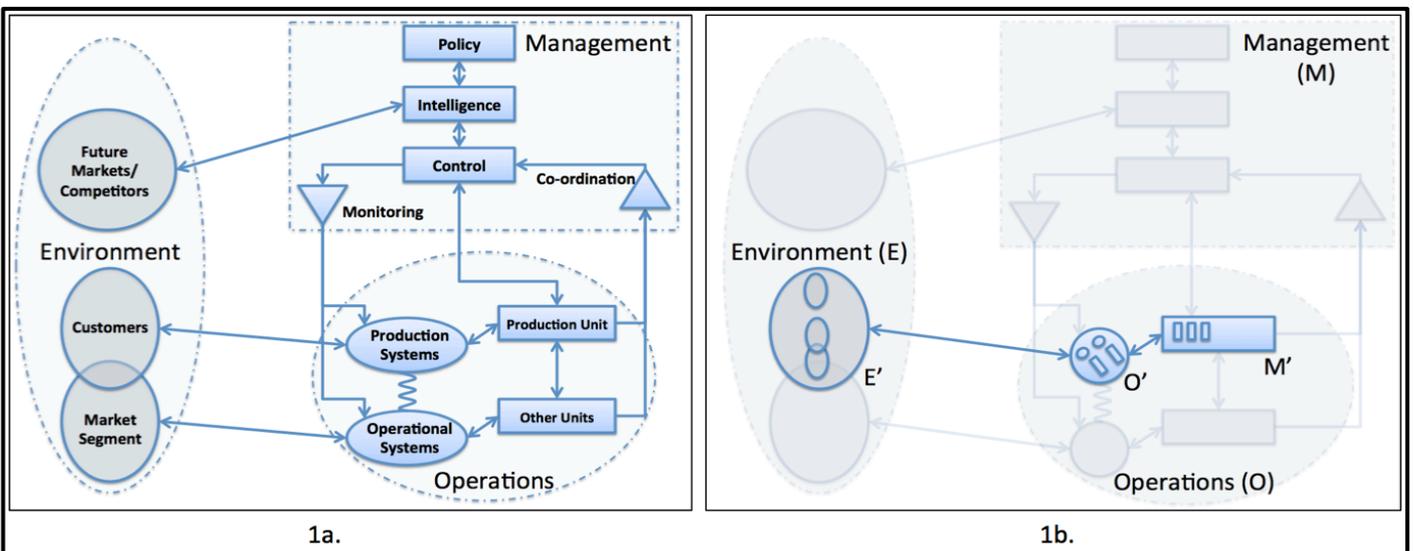

Fig. 1 (a) Typical Viable System Model (VSM); (b) Systemic representation of recursive nature of VSM

place inside the system while the Management part controls operations, ensures stability, facilitates adaptations to the future trends and structures the organizational policies. The Environment entails all the external entities that exchange data with the system.

Another important aspect related to VSM is that of recursivity, which is based on the premise that all autonomous systems are composed of a series of sub-systems, each having self-organizing and self-regulatory characteristics. Each of the sub-systems contain further sub-systems, and so on, right down to the level of the last viable component and at whatever level they occur, are by definition autonomous. All the viable sub-systems contain the capacity to adapt to the changes and deal with the complexity relevant for their autonomous functioning. Fig. 1b depicts a visual representation of the recursive nature of the VSM.

The VSM provides a comprehensive tool to specify information flows, design system strategies and examine missing components. The inherent dynamics of VSM can be used to expose vulnerabilities of an ICS network by modeling cyber-attacks against the various systems of the model. Bill & Matthew [5] exemplified the use of VSM in cyber-security domain, as a framework for analysis of potential vulnerabilities to an organization's information systems. Their work is limited to a single VSM disregarding its recursive nature and does not provide any information regarding the VSM's relationship with the environment as well as other organizations. In another approach [6] the authors have shown the use of VSM to model Information Security Governance by adopting and simulating the standard BS ISO/IEC 27035. Since their research follows a checklist-based approach using the implementation of a security standard to model the VSM, it focuses exclusively on compliance measures and omits information regarding technical protection mechanisms. Also, their work focuses on a typical corporate network and is not specifically tailored for ICS networks. In yet another approach [7], the authors used VSM as a framework for enhancing traditional risk assessment methods of SCADA systems. In this work, only the SCADA server was modeled as VSM for identification of risk agents without the consideration of other ICS components.

Hence, we introduce a recursive model of secure ICS networks by comprehensively modeling the critical cyber-physical systems layer-by-layer in a novel manner. Furthermore, the proposed model takes into consideration the relationships and interactions between the different entities within and outside the ICS infrastructure, and hence provides a holistic view of ICS networks.

## III. PROPOSED VSM FOR ICS NETWORKS

In ICS security a gap exists between academia and industrial practice. Academic research focuses on highly sophisticated attacks and countermeasures, whereas the failure of many real-world deployments is due to a lack of the simplest security best practices. To close this gap, a layered defense-in-depth framework is required, which can simplify the identification of feasible best practices. Generally, operations team always has high availability and reliability in mind, with low downtime mandates and cyber-security is not a significant concern for them. Hence, introduction of security into C&I systems requires tight integration of security and control system expertise to address security risks and needs.

With the introduction of IT controls in industrial processes several best-practice system models have emerged that rely on isolation. In the proposed VSM this concept of isolation is further extended into segregated recursive layers as per the viability of ICS network. The proposed ICS VSM is stratified into 7 levels of recursion labeled from L7 to L1 as follows:

- L7 ICS Network Management
- L6 IT Operation Network
- L5 ICS-DMZ Operations
- L4 Production Operations
- L3 Process Control Network
- L2 Control System Network
- L1 Device Network.

*L7 ICS Network Management*

The L7 layer provides the overall view of ICS enterprise network, addressing all the upper level systems starting from IT Operations Network (L6) to the Field Device Network (L1). Fig. 2 depicts the top level VSM representation of ICS enterprise, which is divided into three parts viz. 'L7 ICS Network Management', 'Operational Units' and 'Controllers & Dependencies' (Environment) based on VSM principles. Each Operational Unit recursively functions as a VSM in itself. Quality control and inspection audits of all Operational Units are coordinated through the Executive Directors whereas, market analysis and competitor assessment are carried out by the Marketing & Planning Directors.

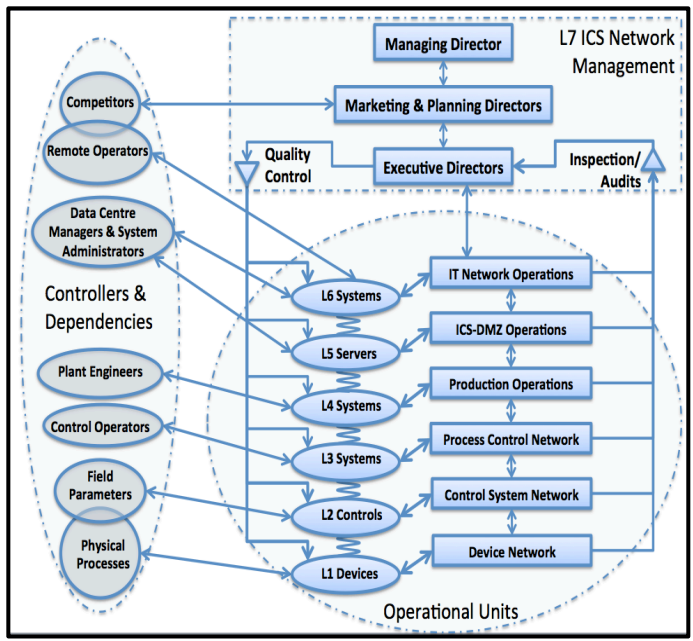

Fig. 2 VSM for L7 ICS Network Management

At first glance, VSM at this top level may seem simple and obvious. However, the lack of an overall well-coordinated system may result in significant effects to the operations in lower layers. For example, a less effective inspection of internal security systems may result in existence of unnoticed critical vulnerabilities, which could have been protected from exploitation if requisite initiatives were taken on time. Thus, even at this level threats that derive from poor organizational management of the ICS enterprise can be identified. For security assurance at this layer, the enterprise must ensure compliance to policies relating to information security standards & must conduct regular audits to inspect and verify security operations in lower layers.

*L1 Device Networks*

The L1 layer represents the VSM of field device network consisting of sensors and actuators for monitoring & control of physical processes. Sensors and actuators are generally wired directly to microprocessor-based controllers called IEDs (Intelligent Electronic Devices). IEDs receive data from sensors and can issue control commands to actuators, such as tripping circuit breakers if they sense voltage, current, or frequency anomalies in order to maintain the desired values. The IEDs are closely linked with real-time systems called PLCs & RTUs for centralized control. PLCs use Fieldbus communication protocols to communicate with array of field devices. RTUs are more suitable for wide geographical telemetry and often use wireless communications. Fig. 3 depicts the VSM representation of the field network, which is divided into three parts viz. 'L1 Device Network' (Management), 'Devices' (Operations) and 'Field Entities' (Environment) as per VSM design principles.

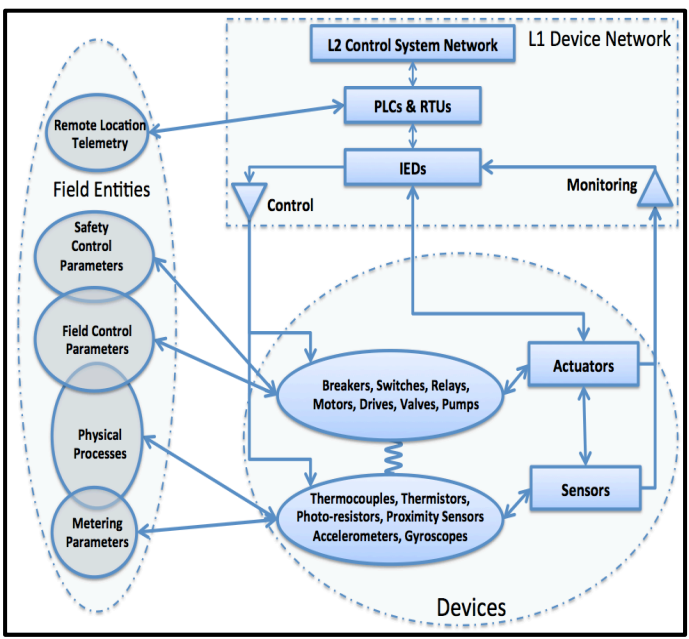

Fig. 3 VSM for L1 Device Network

The impact of cyber-attacks in L1 layer is much higher, because it can directly disrupt physical processes. Generally, field devices apply COTS components to custom boards with processor architectures like ARM, Power, Motorola 68k etc. similar to those in non-ICS embedded systems. Most of these field devices implement only rudimentary security features, such as requiring a passcode to alter configuration, and some do not have any security protections. Evident attacks [8] exploiting these inherent hardware flaws violate the often-made wrong assumption that field devices aren't vulnerable because they are isolated from the business network & Internet. Thus, it is necessary to implement security controls at this essential layer considering the criticality of physical processes. Some security control measures may be like implementing granularity control of bandwidth with restrictions in the rate of flow of information (up to a very low order of few bits per second) sufficient enough to cater transmission of field data.

Also, the more susceptible wireless communication devices can be minimized or barred wherever possible to prevent digital intrusion.

*L2 Control System Network*

The L2 layer represents the VSM of control system network consisting of various control modules like sensor control, drive control, batch control, safety control etc. as shown in Fig. 4. The VSM is divided into three parts viz. 'L2 Control System Network' (Management), 'Control System Modules' (Operations) and 'Control Entities' (Environment). The telemetry data received by PLCs of each control module are deterministically processed in real-time using control logic algorithms and ladder logic sequences stored in PLCs' memory. Most PLCs are programmed using application software running on personal computers called PLC Terminals which are connected to the PLC through Ethernet, RS-232 or RS-485 cabling.

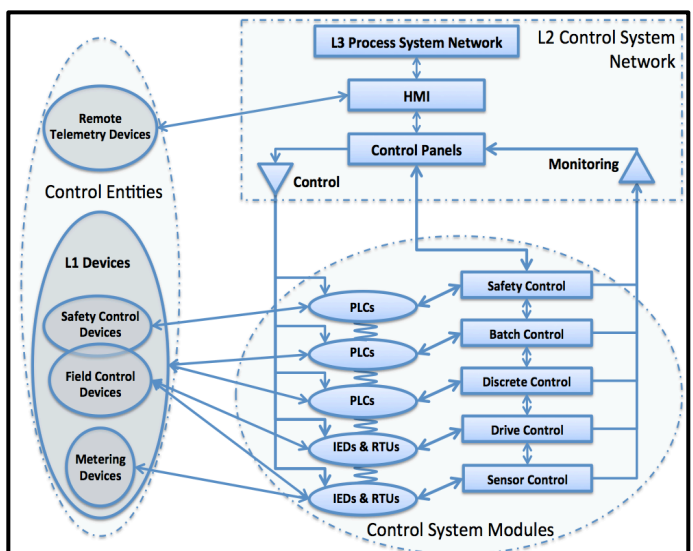

Fig. 4 VSM for L2 Control System Network

The control algorithms running in the PLCs of L2 control modules determine the critical functioning of C&I systems. Most deployed PLCs contain COTS Operating Systems like VxWorks and OS-9 as well as embedded Windows and Linux variants. Vulnerabilities and exploits for these Operating Systems (OSs) exist much as they do for regular desktop OSs such as Microsoft Windows. PLCs can be attacked indirectly by gaining control of terminals that communicate with them. This is what happened in the case of Stuxnet worm. Currently many PLCs accept new logic without any authentication, and often don't allow inspection of their state. Hence, in order to protect control module PLCs, communication and interaction among them should be secure and trustworthy. Trustworthiness can be implemented through the use of software or hardware such Hardware Security Modules (HSM). Any new code update should be authenticated and authorized using cryptographic techniques before execution. Though cryptographic techniques can be resource intensive for PLCs, but if algorithms and protocols are rightly optimized it is worth the security. Integrity checks for executable software can also be enforced at boot up time.

*L3 Process Control Network*

The L3 layer represents the VSM of process control network that is used to transmit instructions to the L2 control network using centralized SCADA systems. Fig. 5 depicts the VSM, which is divided into three parts viz. 'L3 ICS Network Management', 'Process Control Systems' (Operations) and 'Process Controls' (Environment) as per VSM design principles. The L3 layer consists of systems like Batch management systems, Alerting, Alarm systems, Historian Archive systems etc., which interact and communicate with L2 controls. The key requirements for this layer are robustness and determinacy. Robustness involves connection redundancy, reduced sensitivity to Electromagnetic Interference (EMI), good error checking & correction etc. whereas determinacy involves guaranteed network access, high availability of priority systems etc.

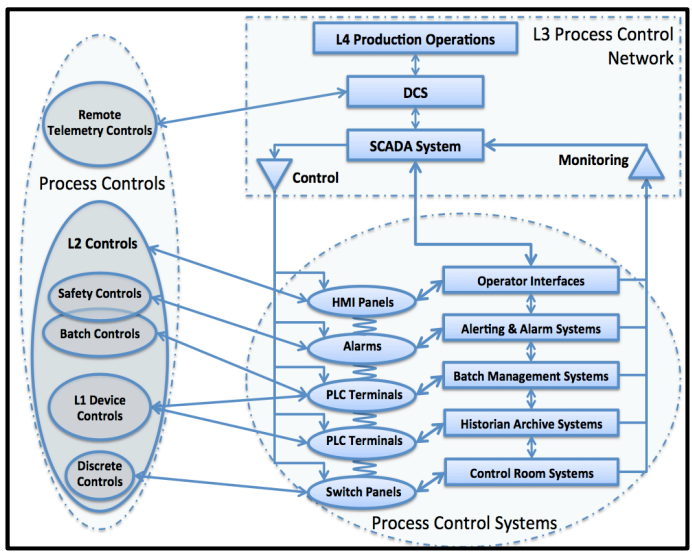

Fig. 5 VSM for L3 Process Control Network

As C&I systems are discontinuing the use of proprietary protocols and adopting new COTS components for increased flexibility, exposure to cyber threats has increased. Many modern data acquisition systems of process control network rely on Ethernet, TCP/IP, and Windows technology instead of old serial communication protocols like Modbus, and hence are more vulnerable due to interoperability with regular IT network. In addition to this, legacy devices and protocols bear a variety of vulnerabilities, which are difficult or even impossible to be patched due to lack/end of support. In many cases, replacement of these legacy systems with new, more secure systems just isn't feasible. Also, it's evident that legacy systems are insecure-by-design and as replacements are not always possible, they must be secured by indirect means. Secure communication technologies like Public Key Infrastructure (PKI) can be established based on electronic certificates, digital signatures and cryptographic keys. Application whitelisting is another security control that may be implemented in this layer. Also, operations team must not assume security just by ensuring obscurity of protocols, isolation of networks, and assuming disinterest in potential attackers. Rather, they must proactively plan for prevention of sophisticated cyber-attacks and spread requisite awareness.

*L4 Production Operations*

The L4 layer represents the VSM of the Production operations of an ICS enterprise like Distributed Control System (DCS) management, Asset management, Quality & Reporting etc. Fig. 6 depicts the VSM, which is divided into three parts viz. 'L4 Production Operations' (Management), 'Production Systems' (Operations) and 'Control Room Systems' (Environment). ICS control rooms or supervisory stations consist of multiple servers, distributed software applications, various electronic knobs, switches, HMI panels, large displays etc. Servers are often configured in a dual-redundant or hot-standby mode for providing high availability (continuous control and monitoring) in case of a server malfunction or hardware failure. The operators interact with SCADA systems through the HMI displays to remotely operate field systems, troubleshoot problems, develop and initiate reports, and perform other operations related to centralized administration of distributed C&I systems.

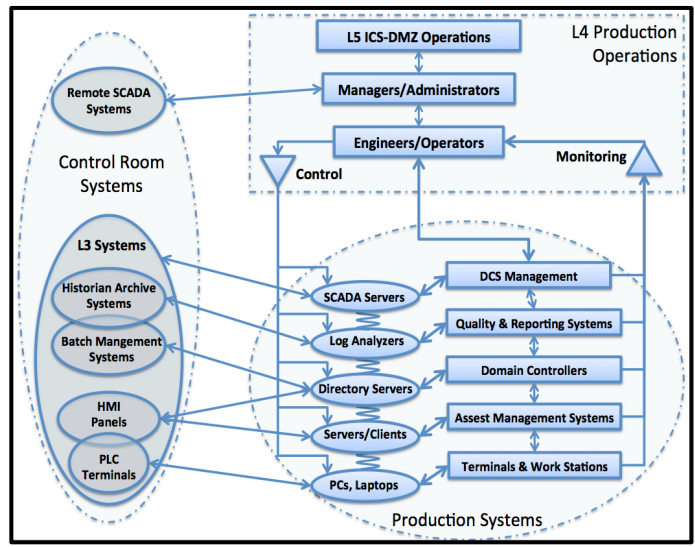

Fig. 6 VSM for L4 Production Operations

The HMI screens provide the easiest method for clear understanding of the underlying C&I processes. Each ICS vendor uses different schema and naming conventions for the SCADA points database in Historian servers, but nearly every C&I system assigns each sensor, pump, breaker, etc., a unique number. On the communications protocol level, the field devices are simply referred to by that unique number. An attacker that gains a foothold on HMI systems directly or indirectly can quite easily discover the point reference numbers and hence exploit the SCADA systems. Generally, L4 production systems are protected by isolating network interfaces with multi degree firewalls into a separate De-Militarized Zone (DMZ) to segregate them from the L3 Process Control Network.

L4 Production systems face a major threat from insiders. Examples include disabling of air conditioning in a data center or allowing chemical processes to occur longer than equipment tolerance levels etc. Insiders may even alter information about key processes at the source of the data to present different information to production systems and hence mislead other operators. Therefore, specific access rights to system administration and documents containing operating procedures, blueprints, ladder logic diagrams, etc. of critical systems should be limited by the principle of least privilege and separation of duties. Log analyzer tools must be used for monitoring of system logins, suspicious behavior patterns, new system connection attempts etc. Such monitoring technologies are also introducing capabilities to integrate with physical security systems. For example, initiating a monitoring sequence to follows the operator's access to specific buildings, rooms and workstations based on RFID badge swipe timings. This access can be automatically validated, logged and then compared against the employee's approved work schedule.

*L5 ICS-DMZ Operations*

The L5 layer represents the VSM of De-Militarized Zone (DMZ) Operations of an ICS datacenter, which isolate and protect the ICS perimeter network from the regular enterprise IT network. Fig. 7 depicts the VSM, which is divided into three parts viz. 'L5 ICS-DMZ Operations' (Management), 'DMZ Servers/Systems' (Operations) and 'Perimeter Network' (Environment). The L5 DMZ Operations include firewall administration, Intrusion detection & prevention (IDS/IPS), Network Access Control (NAC), sandboxing operations, update services & patch management etc. Modern control centers use a combination of Unix, Windows, and web-based management tools for centralized administration and security of L4 production servers.

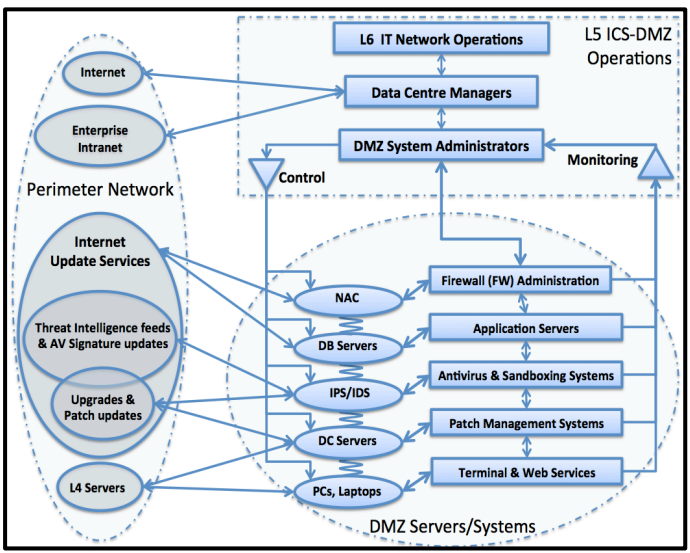

Fig. 7 VSM for L5 ICS-DMZ Operations

Inspectability, the capability to monitor a system's internal state is the backbone of traditional security tools. Most desktop security tools observe the behavior, output, and code signature patterns of system processes. This is how virus scanners identify malware and how integrity checkers identify modifications to important system files. Similarly, network intrusion detection systems rely on the ability to inspect network traffic. Forensics and reverse-engineering tools too require the ability to inspect code and binaries both statically & dynamically as they run on a system. Embedded systems and legacy devices are much less inspectable than desktop computers due to lack of tools and suitable interfaces. Hence, inspection capability is generally possible only in higher VSM levels above L3. Some control system vendors have started addressing this issue by developing external SCADA monitoring systems.

Security administrators use Vulnerability Assessment and Penetration Testing (VA/PT) tools to discover and exploit vulnerabilities. However, the active techniques of port scanning, service fingerprinting, and rapid probing hosts to determine vulnerabilities can negatively impact L4 production servers and L3 Process control systems. If the implementation of communication protocols is not robust because of poorly written network daemons or issues with legacy systems, a simple port scanning may result in slow performance or lead to a crash and subsequent denial of service. Hence, some organizations adopt network policies that forbid traditional VA/PT assessments on production networks. As a solution to this SCADA specific passive assessment tools are coming up, which do not impact production systems and can be automated to run only in non-peak hours. One significant problem these VA/PT tools can't resolve is that of protection of C&I servers from advanced malware exploiting zero-day vulnerabilities. Traditional defense mechanism like Intrusion Detection/ Prevention Systems (IDS/IPS) or Network Antivirus use signature based detection technologies and hence cannot detect zero-day threats. Modern security solutions like Malware Sandboxing systems are coming up which apply dynamic analysis of executable code in a virtual sandbox environment to proactively detect suspicious code behavior and block execution.

Moreover, the threats to L5 systems generally arise due to operational negligence, inappropriate testing, vulnerable end-points, wrongly defined security policies and even false perception of security. Hence, just an assumption that having industrial firewalls and VPN solutions offers sufficient protection, can pose significant threat at this layer. Datacenter managers must review compliance of clearly defined cyber security controls and verify implementation of necessary security patch updates.

*L6 IT Operation Network*

The L6 layer represents the VSM for IT operations of an ICS network. Fig. 8 depicts the VSM, which is divided into three parts viz. 'L6 IT Network Operations' (Management), 'Datacenter Operations' and 'IT Network Resources' (Environment) as per VSM principles. This layer encompasses facility-related components that support enterprise IT operations like ERP, Mail, VPN, Intranet & Internet services etc. along with administration of remote Datacenters and Disaster Recovery sites. The Chief Information Officer (CIO) of the ICS enterprise centrally coordinates quality control and inspection audits of all L6 IT Operations.

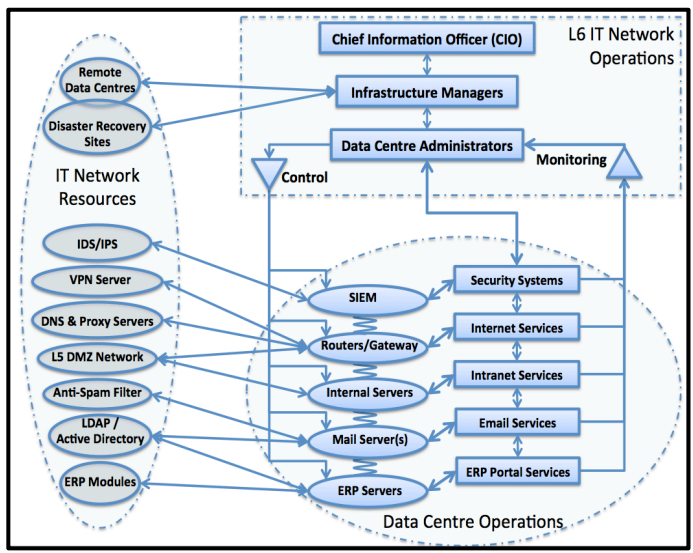

Fig. 8 VSM for L6 IT Network Operations

The L6 IT Operation Network layer is very critical; in the sense, a security breach here provides a gateway to penetrate into lower levels. Also, the L6 systems & services are quite diverse in nature and more risk-prone due to connectivity with Internet. Hence, a plethora of security systems are required in this layer ranging from contemporary antivirus, firewalls, IDS/IPS, secure web/mail gateways to modern malware sandboxing, DLP (Data Loss Prevention) and SIEM (Security Information and Event Management) systems. SIEM systems provide a comprehensive platform to detect, monitor, record and analyze security events, for real-time identification of threats through centralized correlation of logs generated from various network entities like servers, end-user systems, firewalls etc. SIEMs empower security administrators by providing an all-in-one proactive security management solution to correlate events from multiple sources. SIEMs with advanced communication profiling technologies can also be extended to lower layers, which can detect suspicious attempts like sending of critical data through uncommon network paths. Often in L3, L4 and L5 operations, engineers have a detailed view of what their process is doing but no idea of what their network is doing. This is because; current security monitoring tools provide too much data and not enough information. The plant floor is flooded with data and alerts on potential safety, production, maintenance, inventory, and environmental issues. Security information gets lost in the noise. Hence, a good SIEM system must figure out how to get the right security information at the right time without generating false positives and without flooding the dashboard with unnecessary data.

Another important aspect of the L6 layer is disaster management or the recovery of information/system after occurrence of an unavoidable intrusion/malware attack. To address this, virtualization technologies are heavily used for management of IT resources. Virtual Machine (VM) snapshots and clones can shrink-wrap operational systems for fast recovery and provide features like fail-over, High-Availability (HA) and Fault Tolerance (FT). Virtualization also helps in reduction of hardware maintenance & operational costs. Though the penetration of virtualization in lower layers is not as much as in L6 layer, but modern L3 SCADA systems and L4 production systems are adopting virtual infrastructure day by day. However, virtualization has increased the attack surface because of the hypervisor layer in the middle. A compromise of the hypervisor could result in the compromise of all hosted workloads/VMs. Also, the lack of visibility and controls on internal virtual networks created for VM-to-VM communications blinds network-based IPS/IDS systems. Hence, agent-less virtual anti-virus appliances are coming up that provides malware protection to VMs. Since these appliances provide an agent-less approach the processing doesn't burden the individual VMs. The drawback to some of these appliances is that they only provide antivirus protection to VMs and not the additional application control, IDS/IPS and web filtering contained in more traditional appliances. Other security measures in virtualization may be like implementation of MAC address filtering, placing of virtual switches into promiscuous mode for packet inspection by SIEMs, introducing virtual firewalls for VMs in different VLANs etc.

## IV. IMPLEMENTATION INSIGHTS

ICS security requires tight integration of information security and industrial control system expertise. Engineers know that control is impractical if the accuracy of inputs and the reliability of outputs are questionable. Security experts know that security without integration with operations is ineffective. All this is known for years, and may seem obvious, but this relationships, interactions and organizational issues (management, coordination etc.) can't be neglected if security issues are to be addressed. One of the striking features of the proposed ICS VSM is that it is built upon these relationships, which makes it easy adaptable and automatable for industry practitioners to implement it in the real world. The inherent structure of VSM dynamics is highly employee-oriented, with a segregated level of trust and control at different levels. Each managerial entity of the model is provisioned with only necessary & sufficient control authorities based on the principal of least privilege. Similarly, each operational entity is restricted to do only requisite operations that are intended for them with minimum deviation from their original line of work. This tends to equate the adaptation requirements of both security and operational workforce, because it streamlines the best practices of both the worlds. The VSM model clearly demonstrates how the managers at each layer have to identify the demands of security expertise, while maintaining the focus on operations and environment.

Also, the higher the level of VSM, the more stringent is the requirement to balance the demands of different entities and steer the system as a whole. The lessons learnt from the above analysis of each VSM layer reveal that, defense-in-depth security strategies for various entities depend on their positions within the VSMs. This is where the key benefit of implementing ICS VSM lies. It identifies the important areas where security control is needed while depicting the impact of their compromise on other entities as well. This helps security decision makers make the right decision by enabling them to put more emphasis on the critical areas. The ICS VSM brings out the insight that security initiatives at a particular VSM layer not just protects the entities within the layer, but also is responsible for the security of the recursive entities in the layers below & above it. Hence, with VSM as a reference, the decision makers are empowered to adopt a much more graded approach for implementation of security controls.

## V. CONCLUSION

The paper demonstrated the application of principles of VSM to holistically model the critical cyber-physical systems of an ICS infrastructure. The recursive nature of VSM helped in identifying the relationships & interactions among the self-organizing & self-regulatory subsystems of ICS infrastructure. From the context of VSM it is evident that the viability of the whole ICS network is inextricably linked with the viability of its subsystems at different recursion levels and therefore each subsystem functions as a VSM in itself. The VSMs from the L7 ICS management layer to the L1 field device network provided a framework to understand the complexity of C&I systems and helped in the analysis of their susceptibility against modern cyber-attacks. The paper discussed some key security issues in each VSM layer and their respective mitigation measures. A detailed granular security analysis of all practical possibilities at each layer is not addressed in this paper, as the intention was to just introduce the promising potential of VSM as a vehicle to drive ICS security decisions.

The paper also discussed how the implementation of ICS VSM empowers the industry practitioners & security decision makers to successfully identify critical areas of ICS networks and implement requisite security controls at different levels of recursion. The proposed ICS VSM serves as a dynamic tool to specify information flows, identify bottlenecks, design system strategies and examine missing components that hinder security mitigation efforts. The VSM can also be used as input to regular risk management tools. Furthermore, a quantitative risk evaluation process can be developed that can make use of the findings of the VSM to compute risks. As part of future work attribute-based modeling schemes are being evaluated, where each sub-component will embody certain characteristics based on its position within the VSMs and integrate them with decision and game theoretic models to quantify their behavior in different situations.